\documentclass[twocolumn,showpacs,preprintnumbers,amsmath,amssymb]{revtex4}
\usepackage{epsfig}
\usepackage{array}
\newcommand{\LPC}[0]{\it LPC Caen (IN2P3-CNRS/ENSICAEN et Universit\'e),
F-14050 Caen Cedex , France}

\begin{document}

\title{Comparison of central collisions with thermalised systems
in the framework of the classical N-body dynamics.}
\author{F.~Morisseau}
\affiliation{\LPC}
\author{D.~Cussol}
\affiliation{\LPC}

\date{\today}
 
\begin{abstract} In this study we address the question of the thermal description of collisions of classical clusters in the framework 
of classical N-body dynamics. 
We compare the results of  systems in central collisions and those of
thermalised systems with the same  sizes and with the same available energies. The comparisons are made on size distributions, on 
total multiplicities distributions and on IMF $( N>3 )$ multiplicities distributions. These distributions are found identical for 
available energies per particle below the energy of the least bound particle of the total system. 
They are notably different for higher available energies. 
These results may put a new light on the standard explanation for fragment formation in N-body collisions.
\end{abstract}

\maketitle

During the last two decades, strong experimental and theoretical efforts have been put on the interpretation of fragment formation in 
nucleus-nucleus collisions. 
This question is of great importance for the determination of the properties of the infinite nuclear matter. 
Experimentally a strong enhancement of fragment multiplicity with a charge higher or equal to 3 has been observed for excitation 
energies per nucleon greater than to roughly 3-4 MeV. 
The usual way to theoretically describe such an enhancement is to assume a two step process: 
a dynamical stage during which light charged particles are promptly emitted and hot sources are formed; 
and a decay stage during which the hot equilibrated sources decay through thermal emissions of particles and fragments. 
Many codes have been developed to describe these two phases, but the question of whether the fragments are produced during the first 
dynamical stage or during the thermal decay stage is not yet clearly fixed. 
Even if fragment size distributions have been successfully described by statistical fragmentation models, many experimental results 
indicate that the angular distributions and the kinematical properties of the fragments are strongly influenced by the entrance channel.

In this article, we will address the following question in the framework of the Classical N-Body Dynamics code (labeled CNBD): 
can the fragment production be described by the decay of a thermalised system?
We hence have compared the results issued from the collisions of classical systems to the results issued from the decay of thermalised
 systems. In this article, we have limited our comparison to the most central collisions.

First of all we briefly describe the CNBD code and the building of thermalised systems with it. 
Then, we will compare the total multiplicity distributions, the fragment multiplicity distributions and the fragment sizes 
distributions. 
Finally we will draw conclusions.   


The dynamical code CNBD  is described in details in references \cite{CNBD1,CNBD2}.
The evolution of each particle is driven by the Newtonian equations of motion.
The two-body potential used to describe the interaction between the particles is a third degree polynomial.
It has the basic properties of the Lennard-Jones potential \cite{Campi2003,Dorso99}, i.e. it is repulsive at short distance, attractive at medium distance,
and null at long distance.
Since one wants to study the simplest case, neither long range repulsive
interaction nor quantum corrections like a Pauli potential have been introduced \cite{Dorso87}.
A system is simply defined as a set of $N$ particles. When the configuration of these particles minimizes the total energy, the system is considered to be in its ``ground state''.
The dynamical evolutions of the particles of colliding systems and the particles of thermalised systems are described with this code.

In order to avoid any confusion with nuclear physics, the units used here are
arbitrary. The distance will then be in Distance
Simulation Units ($D.S.U.$), the energies in Energy Simulation Units ($E.S.U.$),
the velocities in Velocity Simulation Units ($D.S.U./T.S.U.$) and the reaction time in
Time Simulation Unit ($T.S.U.$).

The ``ground states'' of systems made with CNBD are qualitatively
close to that of nuclei \cite{CNBD1}; and results of collisions between such classical clusters have already been shown in references 
\cite{CNBD1,CNBD2}.
Central collisions are defined as those that have an impact parameter b lower than one tenth of the maximum impact parameter 
$b_{max}$, which is the sum of the radii of the projectile and the target and the range of the two-body potential ($\frac{b}{b_{max}}<0.1$).
For these collisions, the fragment (cluster with at least four particles) multiplicity has been found to strongly 
increase when the available energy per particle is above the energy of the least bound particle $E_{LeastBound}$ of the fused system. 
This energy has also been found to be the above limit of the thermal energy per particle that a free cluster can sustain \cite{CNBD2}.
To perform the comparison of thermalised systems with those in collision, we have build thermalised systems in the following way:  
i) the stable system of interest is put in a sphere with a radius $R_{conf}$. 
Its center of mass coincides with the center of the sphere.
Particles cannot escape this sphere because of a recall force derived
from a quadratic potential.  
ii) The system receives a given amount of kinetic energy. 
iii) Then the particles move in the sphere according to the classical equation of motion
until the system is thermalised. The thermalisation is estimated by looking at the ratio
$\frac{\sigma_c}{<E_c>}$, where $\sigma_c$ and $<E_c>$  are respectively the
dispersion and the mean value of the kinetic energies of the particles.  When this
ratio becomes and then stays close to $\sqrt{\frac{2N-2}{3N-1}}$ with $10^{-2}$ near, the system is considered as thermalised. 
This latter value is the expected value of the former ratio for an isolated system of size $N$ for a classical system in the 
microcanonical ensemble \cite{Lopez89}. 
This procedure is equivalent to the procedures used in other works (see for example \cite{Campi2003,Cherno2004}).

At this stage we could compare directly the results issued from the thermalised systems to those of the systems in collision at the 
freeze-out time which is the time at which the fragments are well separated in the configuration space and do not interact anymore 
with each other. Unfortunately this time is extremely hard to determine for each collision. 
It has been more easier to let the thermalised freely decay during a time corresponding approximately to the difference between the 
ending time of the collision and a very rough determination of the freeze-out time. 
Hence, after the thermalisation time, the confining sphere is removed and the particles of the thermalised systems can move freely 
during roughly 170 T.S.U.. 
This time does not need to be precisely defined since at large collision time, the emission rate is very low and the size distribution 
and the multiplicity distribution evolve very slowly. 

All along this process, the evolution of the classical particles in the system are followed by using the same algorithm as the one which 
is used for describing the systems in collision, and the same two body interaction is used. 
This allow us to keep everything under control and to be sure that the similarities or the differences between these two systems will be 
only due to the different conditions of preparation of the systems.

The comparisons have been performed for systems with 26, 68, 100 and 200 particles. 
The results of a thermalised system with $N$ particles have been compared to the most central collisions ($\frac{b}{b_{max}}<0.1$) of a 
projectile with $N/2$ particles striking a target with $N/2$ particles. 
The same total energy in used in the two cases: the excitation energy of the fused system in the colliding case has been given to the 
thermalised systems. 
The comparisons have been done for four excitation energies values: 
i) an excitation energy per particle $E*/N$ below the binding energy of the least bound particle to the system $E_{LeastBound}$ ; 
ii) an excitation energy per particle close to $E_{LeastBound}$; 
iii) an excitation energy per particle close to the binding energy per particle of the system $E_{Bind}/N$; 
iv) and an excitation energy per particle greater  than $E_{Bind}/N$. 
For the thermalised systems, the influence of the size of the confining sphere has also been checked. 
Three effective densities have been studied: $\rho=\rho_0$ ($R_{conf}=R_{max}(N)$), $\rho=\rho_0 / 3$ 
($R_{conf}=\sqrt[3]{3} R_{max}(N)$) and $\rho=\rho_0 / 8$ ($R_{conf}=2 R_{max}(N)$); $R_{max}(N)$ is the distance between the
centre of mass of the considered system and the particle which is the most distant from it. 
Since for the central collisions one hundred events are selected, one hundred thermalised events have been generated for each set of 
parameters ($E*/N$,$N$,$\rho$).

\begin{figure} 
\begin{center}
\includegraphics[width=3.5in]
{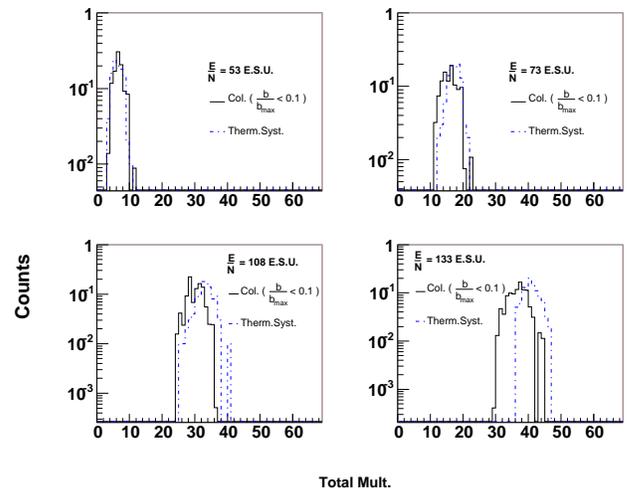}
\caption{Normalized distributions of the total fragment multiplicity for the system $N$ = 68. 
The available energy in the center of mass increases from left to right and from top to
bottom (see text). The full lines correspond to the central collisions of a projectile with 34 particles hitting a target with 34 particles. The dashed lines correspond to the
thermalised systems at normal density ($\rho = \rho_0$). } 
\label{hM} 
\end{center} 
\end{figure}

\begin{figure} 
\begin{center}
\includegraphics[width=3.5in]
{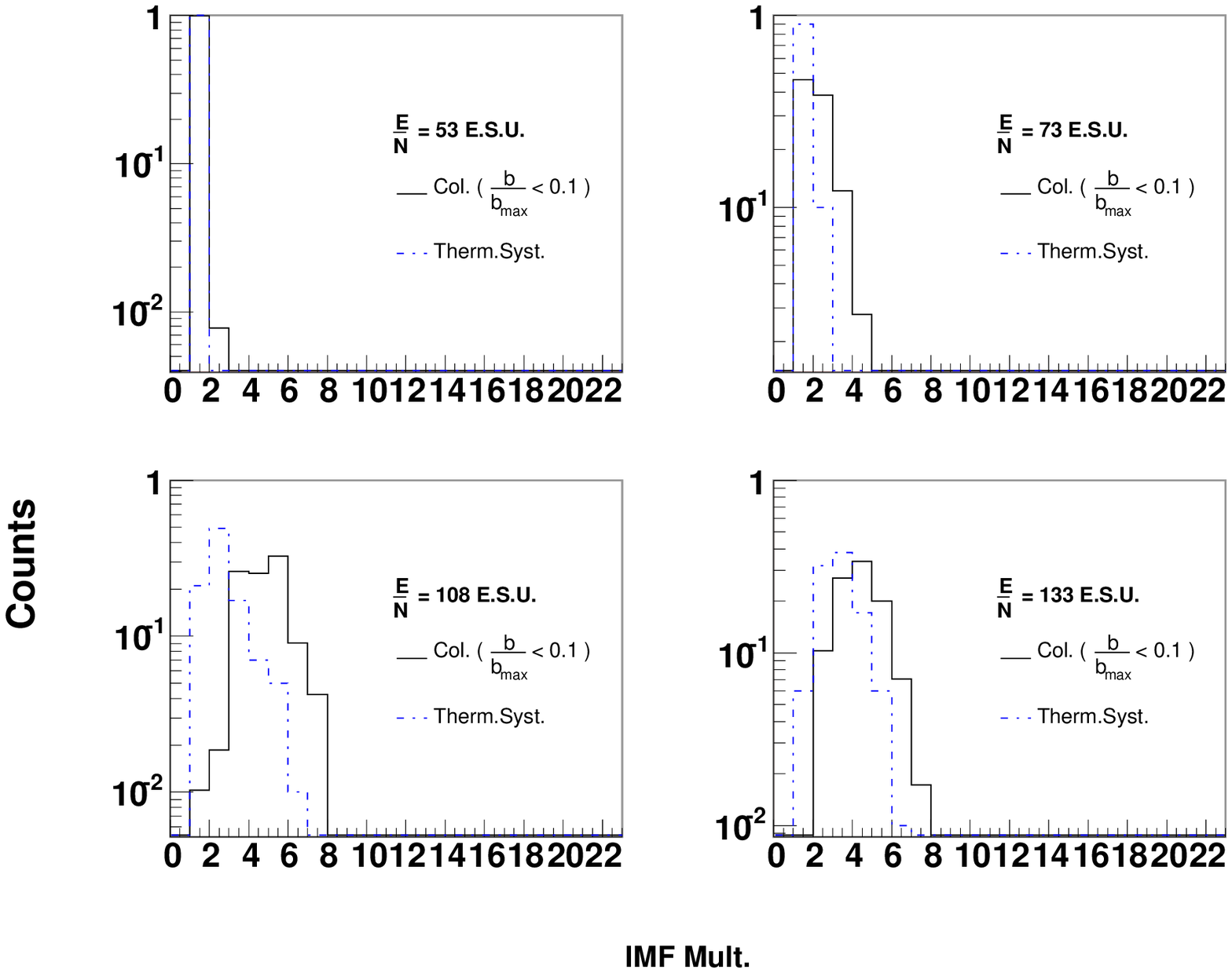}
\caption{Normalized distributions of the IMF multiplicity $( N > 3 )$ for the system $N$ = 68. 
The available energy in the center of mass increases from left to right and from top to
bottom (see text). The full lines correspond to the central collisions of a projectile with 34 particles hitting a target with 34 particles. The dashed lines correspond to the
thermalised systems at normal density ($\rho = \rho_0$). }
\label{hMIMF} 
\end{center} 
\end{figure}

\begin{figure} 
\begin{center}
\includegraphics[width=3.5in]
{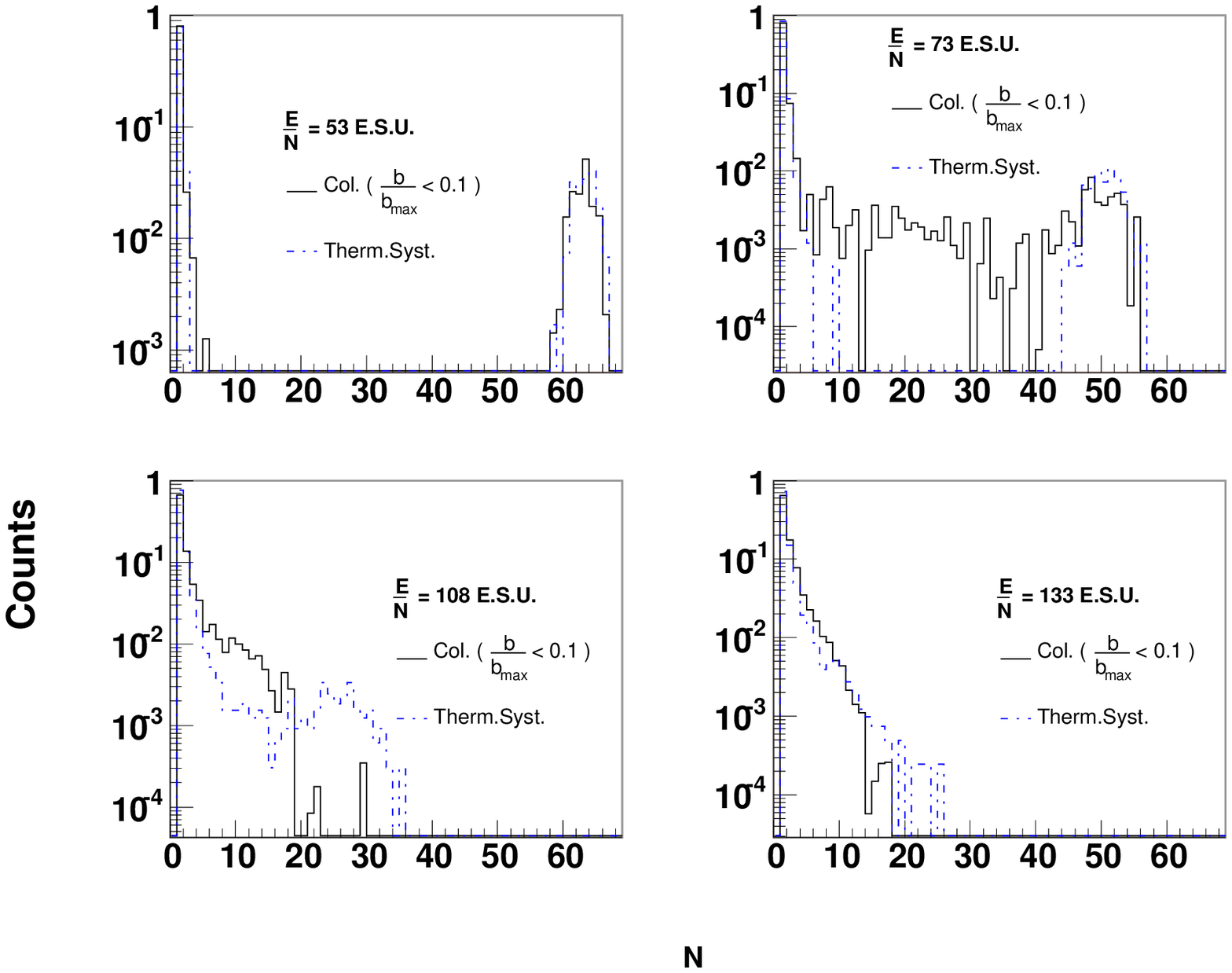}
\caption{Normalized distributions of the sizes of the final clusters for the system $N$ = 68. 
The available energy in the center of mass increases from left to right and from top to
bottom (see text). The full lines correspond to the central collisions of a projectile with 34 particles hitting a target with 34 particles. The dashed lines correspond to the
thermalised systems at normal density ($\rho = \rho_0$). } 
\label{hN} 
\end{center} 
\end{figure}

\begin{table*}
\begin{center}
\begin{tabular}{c c | c c c c c c}

\hline
\hline
$N$&$E^*/N$ &
\multicolumn{2}{c}{Total Multiplicity}&
\multicolumn{2}{c}{IMF Multiplicity}&
\multicolumn{2}{c}{$N$ of the biggest cluster}\\

($E_{LeastBound}$)&(E.S.U.)&\makebox[2cm]{C} &\makebox[2cm]{T} &\makebox[2cm]{C}&\makebox[2cm]{T}&\makebox[2cm]{C} &\makebox[2cm]{T} \\
\hline
& 37 & 1.7 $\pm$ 0.6 & 1.6 $\pm$ 0.6 & 1 $\pm$ 0 & 1 $\pm$ 0 & 25.3 $\pm$ 0.6 & 25.4 $\pm$ 0.6\\ 
26& 60 & 7.2 $\pm$ 1.0 & 6.8 $\pm$ 1.3 &  1.4 $\pm$ 0.5 & 1 $\pm$ 0 & 17.0 $\pm$ 2.9 & 19.6 $\pm$ 1.4\\ 
(50 E.S.U.)& 87 & 12.0 $\pm$ 1.3 & 12.4 $\pm$ 1.4 & 2.1 $\pm$ 0.8 & 1.5 $\pm$ 0.6 & 7.5 $\pm$ 2.0 & 10.1 $\pm$ 2.9\\ 
& 107 & 14.7 $\pm$ 1.9 & 15.4 $\pm$ 1.7 & 1.2 $\pm$ 0.7 & 1.2 $\pm$ 0.6 & 4.9 $\pm$ 1.4 & 6.1 $\pm$ 2.1\\ 
\hline
& 53 & 6.2 $\pm$ 1.5 & 5.9 $\pm$ 1.5 & 1 $\pm$ 0.1 & 1 $\pm$ 0 & 62.5 $\pm$ 1.6 & 62.9 $\pm$ 1.6\\ 
68& 73 & 15.4 $\pm$ 2.3 & 16.8 $\pm$ 1.9 & 1.7 $\pm$ 0.8 & 1.1 $\pm$ 0.3 & 43.7 $\pm$ 9.3 & 50.2 $\pm$ 2.3\\ 
(60 E.S.U.)& 108 & 29.6 $\pm$ 2.4 & 32.6 $\pm$ 2.6 & 4.3 $\pm$ 1.2 & 2.3 $\pm$ 1.1 & 13.8 $\pm$ 3.1 & 23.3 $\pm$ 5.5\\ 
& 133 & 36.3 $\pm$ 2.9 & 40.4 $\pm$ 2.4 & 3.9 $\pm$ 1.1 & 2.9 $\pm$ 1.0 & 8.7 $\pm$ 2.3 & 10.7 $\pm$ 3.9\\ 
\hline
& 42 & 2.7$\pm$ 1.1 & 2.5 $\pm$ 1.1 & 1 $\pm$ 0 & 1 $\pm$ 0 & 98.3 $\pm$ 1.2 & 98.5 $\pm$ 1.2\\ 
100& 72 & 18.6 $\pm$ 2.9 & 20.5 $\pm$ 2.2 & 1.5 $\pm$ 0.7 & 1.0 $\pm$ 0.2 & 73.1 $\pm$ 11.9 & 78.5 $\pm$ 2.6\\ 
(65 E.S.U.)& 102 & 35.1 $\pm$ 3.4 & 41.3 $\pm$ 3.0 & 5.0 $\pm$ 1.0 & 2.4 $\pm$ 1.1 & 23.4 $\pm$ 4.4 & 43.4 $\pm$ 7.5\\ 
& 132 & 48.1 $\pm$ 3.6 & 54.2 $\pm$ 3.2 & 6.1 $\pm$ 1.2 & 4.2 $\pm$ 1.5 & 13.0 $\pm$ 3.6 & 18.2 $\pm$ 6.1\\ 
\hline
& 42 & 2.3 $\pm$ 1.1 & 2.3 $\pm$ 1.1 & 1 $\pm$ 0 & 1 $\pm$ 0 & 198.7 $\pm$ 1.1 & 198.7 $\pm$ 1.1\\ 
200& 72 & 24.6 $\pm$ 3.4 & 27.6 $\pm$ 3.0 & 1.1 $\pm$ 0.3 & 1.1 $\pm$ 0.3 & 174.1 $\pm$ 4.0 & 171.1 $\pm$ 3.7\\ 
(65 E.S.U.)& 102 & 54.2 $\pm$ 4.6 & 66.3 $\pm$ 4.1 & 5.8 $\pm$ 1.1 & 2.3 $\pm$ 1.0 & 49.9 $\pm$ 9.2 & 116.9 $\pm$ 8.6\\ 
& 132 & 81.5 $\pm$ 4.8 & 94.7 $\pm$ 3.7 & 10.2 $\pm$ 1.9 & 6.2 $\pm$ 2.1 & 24.5 $\pm$ 4.7 & 46.4 $\pm$ 14.6\\ 
\hline
\hline


\end{tabular}

\caption{Mean values of the total multiplicity ,of the IMF ($N>3$) multiplicity, and of the size of the biggest  cluster for both
central collisions (C) and thermalised systems (T) for different system sizes.  The
standard deviations $(\sigma)$ of these observables are indicated after the $\pm$ symbol. 
The density of the thermalised systems is equal to $\rho_0$. \label{recap}}

\end{center}
\end{table*}


The distributions of the total multiplicity, of the intermediate mass fragments multiplicity, and of the sizes of the final clusters 
are shown on figure \ref{hM}, \ref{hMIMF} and \ref{hN} respectively. 
The intermediate mass fragments (labeled IMF) are defined as the fragments with at least four particles ($N>3$). 
Theses distributions are shown for the system with $N=68$ particles and at the four excitation energies. 
The dashed lines correspond to the thermalised case and the full lines to the central collision case. 
On each figure, the distributions corresponding to the low excitation energy ($E*/N < E_{LeastBound}$) are displayed on the top left 
panel, those corresponding to $E*/N$ close to $E_{LeastBound}$ are displayed on the top right panel, 
the distributions corresponding to $E*/N$ close to $E_{Bind}/N$ are displayed on the bottom left panel 
and those corresponding to the high excitation energies ($E*/N > E_{Bind}/N$) are displayed on the bottom right panel.

On figure \ref{hM} one can clearly see that at the low excitation energy (top left panel), the total fragment multiplicity
distributions for the central collisions and for the thermalised case are almost identical. 
For the highest energies, these distributions differ slightly  at $E*/N\approx E_{LeastBound}$ (top right panel) and at 
$E*/N \approx E_{Bind}/N$ (bottom left panel), the average multiplicity value obtained for the thermalised systems is greater than the 
average multiplicity obtained for central collisions. 
At the highest excitation energy (bottom right panel), the two distributions are clearly different: the thermalised systems emit much 
more fragments than the central collisions.
In the same way, the IMF multiplicity distributions (figure \ref{hMIMF}) are only similar at low excitation energy for the central 
collisions and the thermalised systems. 
For the two intermediate energies ($E*/N \approx E_{LeastBound}$ and $E*/N \approx E_{Bind}/N$) the thermalised systems emit less IMF 
than the central collisions. 
This difference is slightly reduced at the highest excitation energy but the thermalised systems still produce less IMFs than the 
central collisions.

As for the total multiplicity distributions and the IMF multiplicity distributions, the cluster size distributions (figure \ref{hN}) for 
the thermalised systems and for the central collisions are almost identical for the lowest excitation energy. 
But they are very different at the two intermediate energies ($E*/N \approx E_{LeastBound}$ and $E*/N \approx E_{Bind}/N$): 
while  for the central collisions  the intermediate mass fragment area is well populated and no ``fusion residue'' can be clearly seen , the IMF area is empty at $E*/N \approx E_{LeastBound}$ and is barely populated at $E*/N \approx E_{Bind}/N$ 
for the thermalised systems. 
A clear contribution of a ''fusion residue'' is also seen at these two energies for the thermalised systems. 
At the highest excitation energy, the two size distributions seem to be closer to each other, but the slopes of the distributions are 
different and the sizes of the biggest fragments are higher for the thermalised systems.\\

From these distributions one can define two energy domains: one corresponding to $E*/N$ values lower  than $E_{LeastBound}$ where the 
results from the thermalised systems and from the colliding systems are almost identical, and one corresponding to $E*/N$ values greater than 
$E_{LeastBound}$  for which the distributions resulting for the two cases are clearly different from each other: 
while the thermalised systems emit mainly small fragments and keep a ``residue'', the central collisions produce mainly IMFs and no 
more ``residue'' is seen. 
This observation has to be linked to the conclusion made in the reference \cite{CNBD2}: 
the energy per particle that can be stored in a free cluster can not exceed the energy of its least bound particle. 
It is hence not surprising that the thermalised scenario does not correspond anymore to the central collisions when the excitation energy 
per nucleon of the fused system is greater than $E_{LeastBound}$. 
Another conclusion is that the thermalised systems can not produce a large amount of IMFs. 
To form an IMF, the particles have to be close in phase space. 
Such correlations are of course less probable in a completely thermalised case where the whole phase space is uniformly covered than 
in the dynamical case where only a small part of the available phase space is populated, due to the entrance channel effects.
 
These observations seem to be independent of the system size. 
The table \ref{recap} summarizes the results obtained for the other systems sizes. 
The lines correspond to different system sizes. 
In each cell, the average value of the observable of the corresponding column is indicated and is followed by its standard deviation. 
On the top of the column, the $C$ letter corresponds to the central collisions and the $T$ letter to the thermalised systems. 
The two energy domains can be clearly seen: the low energy domain ($E*/N < E_{LeastBound}$) where the results from the thermalised 
systems and the results from the central collisions coincide, and the high energy domain ($E*/N > E_{LeastBound}$) for which the results from the 
thermalised systems and the results from the central collisions differ significantly.


This study shows that in CNBD the description of the central collisions as thermalised systems depends on the excitation energy per 
particle of the fused system. 
If this energy is lower than the binding energy of the least bound particle ($E_{LeastBound}$) of the fused system, the sizes 
distributions, the multiplicity distributions and the IMF multiplicity distributions issued from central collisions are identical to 
those of the thermalised systems with the same sizes and with the same excitation energies per particle. 
The thermalised systems decay through an evaporative process which produces many small clusters and a big  ``evaporation residue''. 
This is consistent with the observation of a fusion/evaporation process observed at low energies in CNBD \cite{CNBD1}. 
This behavior is qualitatively very similar to the behavior of Gemini code \cite{GEMINI} which has been widely used to describe the 
thermal decay of hot nuclei. 
When the excitation energy per particle is greater than $E_{LeastBound}$,
the size distributions, the multiplicity distributions and the IMF multiplicity distributions of systems in central collisions are 
different from those obtained for the corresponding thermalised systems. 
While thermalised systems still produce mainly small clusters and an ``evaporation residue'', a large amount of IMFs are produced 
in central collisions and no more ``residue'' is seen.

It is worth to notice that $E_{LeastBound}$ seems to be a ``key'' energy for these classical systems as in references 
\cite{CNBD1,CNBD2}: it is the fragmentation threshold for colliding classical systems and the upper limit for the energy per particle 
that can be stored in free clusters. 
This predominance of $E_{LeastBound}$, which can also be considered as a surface energy, shows that finite size effects play a major 
role for these classical clusters and can not be ignored for a full understanding of the reaction mechanisms of classical systems.
 
Before extending these conclusions to nucleus-nucleus collisions, one has first to check the influence of a Coulomb-like long range 
repulsive interaction on these results. 
The main deviation to these results may also come from the quantum mechanics. 
This last study is nowadays unfortunately out of reach of such numerical simulations. 
One has also to remind that only central collisions have been studied in this paper and that for peripheral reactions possible 
geometrical effects could alter the observations made here. 
This will be investigated in a forthcoming article.
Nevertheless, providing that Coulomb-like effects and quantum mechanics do not qualitatively alter the conclusions made for these 
classical systems, this could shed a new light on the origins of the multifragmentation process in nucleus-nucleus collisions. 
Additionally, the exact meaning of the thermal description of the fragment  production at high excitation energies 
(typically greater than 4-5 A.MeV) should be reconsidered. 
This could also lead to the re-interpretation of the liquid-gas phase transition signals seen in nucleus-nucleus collisions, 
like the negative heat capacity \cite{Chomaz99,DAgostino2000,DAgostino2002}, 
the bimodality \cite{Chomaz2001,Bellaize2002,Borderie2002,Pichon2004} 
and the Fisher's scaling \cite{Elliott2002}. 
The occurence of such signals in the Classical N-Body Dynamics framework will be studied in forthcoming papers.


\end{document}